# Generation of elliptically polarized nitrogen ion laser fields using two-color femtosecond laser pulses

Ziting Li,[1,2,3] Bin Zeng,[2,5] Wei Chu,[2] Hongqiang Xie,[2,3] Jinping Yao,[2] Guihua Li,[2] Lingling Qiao,[2] Zhanshan Wang,[1,6] and Ya Cheng [2,4,*]

[1] *School of Physics Science and Engineering, Tongji University, Shanghai 200092, China*
[2] *State Key Laboratory of High Field Laser Physics, Shanghai Institute of Optics and Fine Mechanics, Chinese Academy of Sciences, P.O. Box 800-211, Shanghai 201800, China*
[3] *University of Chinese Academy of Science, Beijing 100049, China*
[4] *Collaborative Innovation Center of Extreme Optics, Shanxi University, Taiyuan, Shanxi 030006, China*
[5] *bzeng@foxmail.com*
[6] *wangzs@tongji.edu.cn*
[*] *ya.cheng@siom.ac.cn*

**Abstract:** We experimentally investigate generation of nitrogen molecular ion ($N_2^+$) lasers with two femtosecond laser pulses at different wavelengths. The first pulse serves as the pump which ionizes the nitrogen molecules and excites the molecular ions to excited electronic states. The second pulse serves as the probe which leads to stimulated emission from the excited molecular ions. We observe that changing the angle between the polarization directions of the two pulses gives rise to elliptically polarized $N_2^+$ laser fields, which is interpreted as a result of strong birefringence of the gain medium near the wavelengths of the $N_2^+$ laser.

**References and links**

1. A. Dogariu, J. B. Michael, M. O. Scully, and R. B. Miles, "High-Gain Backward Lasing in Air," Science **331**(6016), 442-445 (2011).
2. A. Laurain, M. Scheller, and P. Polynkin, "Low-Threshold Bidirectional Air Lasing," Phys. Rev. Lett. **113**(25), 253901 (2014).
3. J. Yao, B. Zeng, H. Xu, G. Li, W. Chu, J. Ni, H. Zhang, S.L. Chin, Y. Cheng, and Z. Xu, "High-brightness switchable multiwavelength remote laser in air," Phys. Rev. A **84**(5), 051802 (2011).
4. J. Yao, G. Li, C. Jing, B. Zeng, W. Chu, J. Ni, H. Zhang, H. Xie, C. Zhang, H. Li, H. Xu, S.L. Chin, Y. Cheng, and Z. Xu, "Remote creation of coherent emissions in air with two-color ultrafast laser pulses," New J. Phys. **15**(2), 023046 (2013).
5. D. Kartashov, S. Ališauskas, G. Andriukaitis, A. Pugžlys, M. Shneider, A. Zheltikov, S.L. Chin, and A. Baltuška, "Free-space nitrogen gas laser driven by a femtosecond filament," Phys. Rev. A **86**(3), 033831(2012).
6. S. Mitryukovskiy, Y. Liu, P. Ding, A. Houard, and A. Mysyrowicz, "Backward stimulated radiation from filaments in nitrogen gas and air pumped by circularly polarized 800 nm femtosecond laser pulses," Opt. Express **22**(11), 12750-12759 (2014).
7. J. Ni, W. Chu, H. Zhang, B. Zeng, J. Yao, L. Qiao, G. Li, C. Jing, H. Xie, H. Xu, Y. Cheng, and Z. Xu, "Impulsive rotational Raman scattering of $N_2$ by a remote "air laser" in femtosecond laser filament," Opt. Lett. **39**(8), 2250-2253 (2014).
8. P. R. Hemmer, R. B. Miles, P. Polynkin, T. Siebert, A. V. Sokolov, P. Sprangle, and M. O. Scully, "Standoff spectroscopy via remote generation of a backward-propagating laser beam," Proc. Natl. Acad. Sci. U.S.A. **108**(8), 3130-3134 (2011).
9. J. Yao, H. Xie, B. Zeng, W. Chu, G. Li, J. Ni, H. Zhang, C. Jing, C. Zhang, H. Xu, Y. Cheng, and Z. Xu, "Gain dynamics of a free-space nitrogen laser pumped by circularly polarized femtosecond laser pulses," Opt. Express **22**(16), 19005-19013 (2014).
10. P. Ding, S. Mitryukovskiy, A. Houard, E. Oliva, A. Couairon, A. Mysyrowicz, and Y. Liu, "Backward Lasing of Air plasma pumped by Circularly polarized femtosecond pulses for the saKe of remote sensing," Opt. Express **22**(24), 29964-29977 (2014).

11. S. Mitryukovskiy, Y. Liu, P. Ding, A. Houard, A. Couairon, and A. Mysyrowicz, "Plasma Luminescence from Femtosecond Filaments in Air: Evidence for Impact Excitation with Circularly Polarized Light Pulses," Phys. Rev. Lett. **114**(6), 063003 (2015).
12. J. Yao, S. Jiang, W. Chu, B. Zeng, C. Wu, R. Lu, Z. Li, H. Xie, G. Li, C. Yu, Z. Wang, H. Jiang, Q. Gong, and Y. Cheng, "Population Redistribution among Multiple Electronic States of Molecular Nitrogen molecular ions in Strong Laser Fields," arXiv:1506.03348v2.
13. A. Baltuska and D. Kartashov, in Research in Optical Sciences, OSA Technical Digest (online) (Optical Society of America, Messe Berlin, Berlin, 2014), p. HTh4B.5.
14. Y. Liu, P. Ding, G. Lambert, A. Houard, V. Tikhonchuk, and A. Mysyrowicz, "Recollision induced superradiance of ionized nitrogen molecules," Phys. Rev. Lett. **115**(13), 133203 (2015).
15. B. Zeng, W. Chu, G. Li, J. Yao, H. Zhang, J. Ni, C. Jing, H. Xie, and Y. Cheng, "Real-time observation of dynamics in rotational molecular wave packets by use of air-laser spectroscopy," Phys. Rev. A **89**(4), 042508 (2014).
16. Robert W. Boyd, *Nonlinear Optics* (Academic, 2008).

## 1. Introduction

Recently, lasing action produced in population-inverted assembles, such as nitrogen/oxygen atoms [1, 2], molecular nitrogen ions [3, 4] and neutral nitrogen molecules [5, 6], has attracted a great deal of research interests because of its promising potentials in remote sensing applications [7, 8]. Among them, the mechanism behind the establishment of the population inversion in neutral nitrogen molecules has been well identified, which is due to the impact excitation of neutral nitrogen molecules from the ground state to the $C^3\Pi_u^+$ state by energetic free electrons produced in the intense laser fields [9-11]. However, a widely accepted model accounting for the lasing action in nitrogen molecular ions ($N_2^+$) is still lacking. Yao *et al.* proposed the seed-amplification model and verified the population inversion between the $B^2\Sigma_u^+$ state and the $X^2\Sigma_g^+$ state of $N_2^+$ via energy amplification of a time-delayed seed pulse [3], which was further described as a result of the couplings of the ground and two excited states of $N_2^+$ in the strong laser fields [12]. Kartashov *et al.* proposed that the different rotational periods of aligned molecular ions on the ground and excited electronic states can lead to transient laser gain and thus the creation of the coherent emissions [13]. Liu *et al.* considered the coherent emission as a super-radiant emission whose population transferred to the excited state is caused by the field-induced multiple recollisions [14]. These efforts have significantly enhanced the understanding of the physics of tunnel-ionized molecules in intense laser fields.

In this work, we report on another unusual behavior of the $N_2^+$ laser at the wavelength of 391 nm. Previous results show that the $N_2^+$ lasers induced by tunnel ionization possess the same polarization direction as that of the seed pulses when the pump and seed pulses are either parallelly or perpendicularly polarized to each other [4]. This can be well understood from a seed-amplification point of view. In such a case, the laser signal generated by the seed-amplification mechanism typically inherits all the characteristics of the seed pulses. Interestingly, when the angle between the polarization directions of the two linearly polarized pump and seed pulses is variable in the range of 0°-90°, we find that the $N_2^+$ laser becomes elliptically polarized with a variable ellipticity depending on the angle between the polarization directions of the two pulses. Moreover, the *P*-branch and *R*-branch lines in the $N_2^+$ laser show dramatically different behaviors with the varying polarization direction of the seed pulses (i.e., the polarization direction of the pump is always fixed). We attempt to provide a plausible explanation to qualitatively understand this unexpected observation.

## 2. Experimental setup

The experiment setup is illustrated in Fig. 1(a), which is similar to our previous pump-probe measurement [15]. In brief, a linearly polarized femtosecond laser beam (1 kHz, 800 nm, ~40 fs) from a commercial Ti:sapphire laser system (Legend Elite-Duo, Coherent Inc.) was divided into two with a beam splitter (BS). The first one with a pulse energy of 2 mJ was used as the pump to ionize the nitrogen molecules and build up the population inversion between the $B^2\Sigma_u^+$ state and the $X^2\Sigma_g^+$ state of $N_2^+$. The other beam, after being frequency-doubled by a 0.2-mm thick β-barium-borate (BBO) crystal, was used as the seed to generate the $N_2^+$ laser. A half wave-plate (HWP) was employed to change the polarization direction of the seed pulses. A Glan-Taylor prism (GT1) was inserted before the HWP to ensure that the seed pulses were linearly polarized. The pump and seed pulses were combined by a dichroic mirror (GM2) and focused by an $f$ = 30 cm fused-silica lens into a vacuum chamber filled with nitrogen gas. The time delay between the pump and seed pulses was controlled by a motorized linear translation stage with a temporal resolution of ∼16.7 fs.

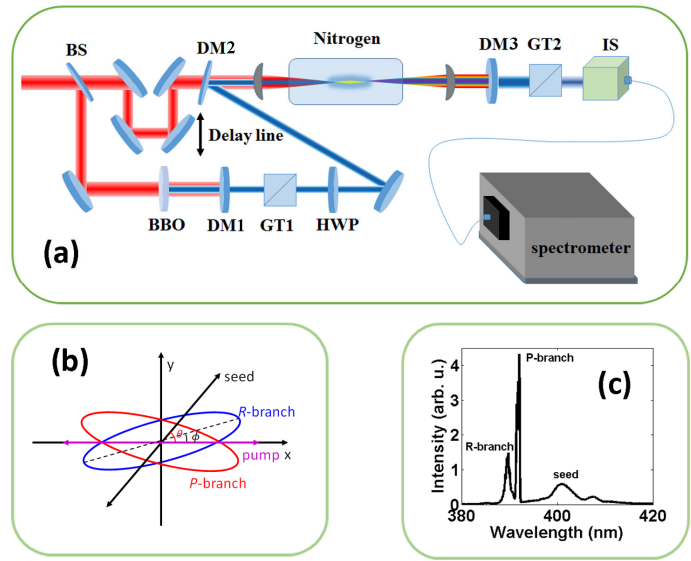

Fig. 1. (a) Schematic of the experimental setup. (b) Polarization states of the pump pulse, the seed pulse, and the P-branch (red ellipse) and R-branch (blue ellipse) lines of the $N_2^+$ laser.
(c) A typical spectrum of the $N_2^+$ laser generated based on the seed-amplification scheme.

The generated $N_2^+$ laser was first collimated by an f = 30 cm lens and then passed through a dichroic mirror (DM3) to filter out the residual pump pulses. To minimize the measurement error caused by a possible spatial anisotropy of the laser and a polarization dependent response of the spectrometer, we used an integral sphere (IS) to collect the signals and directed them into a grating spectrometer (Andor, Shamrock 303i). Another Glan-Taylor prism (GT2) was placed before the IS to measure the polarization of the laser pulses. Figure 1(c) shows a typical spectrum of the $N_2^+$ laser at 391-nm, which is assigned to the first negative band system $(B^2\Sigma_u^+ \to X^2\Sigma_g^+)$ of $N_2^+$. The rotational P-branch $(|J\rangle \to |J+1\rangle)$ and the R-branch $(|J\rangle \to |J-1\rangle)$ lines of the laser are labeled in the spectrum.

Throughout the experiment, we fixed the polarization direction of the pump pulses. We tuned the polarization direction of the seed pulses by rotating the HWP, and measured the intensity of the $N_2^+$ laser as a function of the angle of GT2. The zero degree of the angle of GT2 corresponds to that the optical axis of GT2 is parallel to the polarization direction of the pump pulses.

## 3. Experimental results

Figure 2 shows the intensities of the *P*-branch (red diamonds) and *R*-branch (blue circles) lines of the $N_2^+$ laser as well as the intensity of the seed pulses (black stars) measured after the GT2 as functions of the angle of GT2. The pressure of the nitrogen gas was 13.5 mbar and the time delay between the pump and seed pulses was set at 1.5 ps. The angle θ in the different panels of Fig. 2 are $0°, 10°, 20°, 30°, 40°, 50°, 60°, 70°$ and $80°$. It is noticed that when the angles $θ < 70°$, the $N_2^+$ laser is nearly linearly polarized, whereas its polarization direction is almost parallel to that of the *pump* pulses but not to that of the *seed* pulses. As the angle θ increases to $80°$, the $N_2^+$ laser becomes significantly elliptically polarized with an ellipticity of 0.28. Interestingly, the curves of *P*-branch and *R*-branch lines in Fig. 2 does not overlap, indicating their different polarization characteristics despite their wavelengths so close to each other.

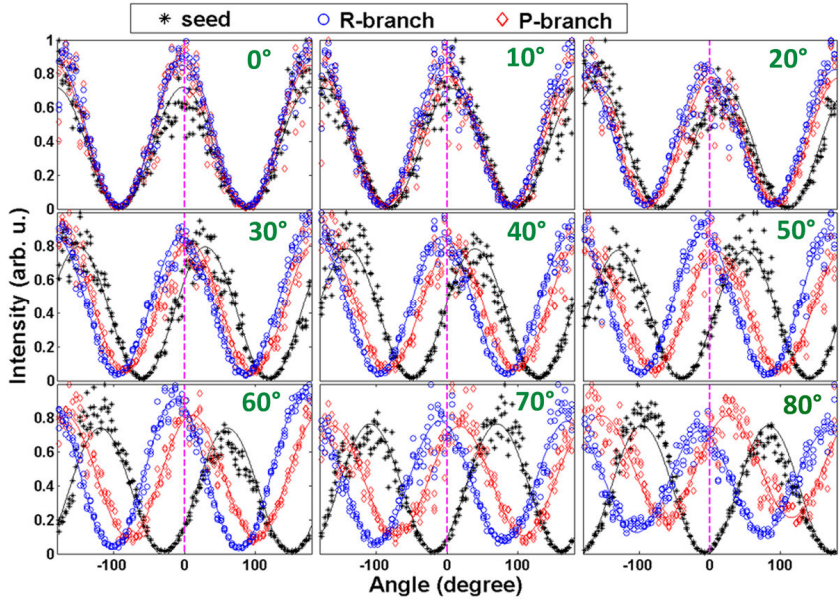

Fig. 2. Measured intensities of the *P*-branch (red diamonds) and *R*-branch (blue circles) lines of the $N_2^+$ laser and the seed pulses (black stars) as functions of the angle of GT2. The angle between the polarization directions of the pump and seed pulses are indicated in each panel. The magenta dashed lines indicate the zero angle of GT2.

For clarity, we plot the azimuthal angle $\phi$ of the $N_2^+$ laser generated at a time delay of ~1.5 ps as the functions of the angle of GT2 (i.e., θ) in Fig. 3. Here, the azimuthal angle $\phi$ is defined as the angle between the major axis of the $N_2^+$ laser and the polarization direction of the pump pulses, as indicated in Fig. 1(b). The solid diamonds represent the azimuthal angle for the *P*-branch laser lines, and the solid circles stand for the *R*-branch laser lines. It is found that the azimuthal angle of the *P*-branch lines are always positive, whereas the azimuthal angle of the *R*-branch lines are always negative. The absolute values of the azimuthal angles of both the *P*-branch and *R*-branch lines increase with the angle θ.

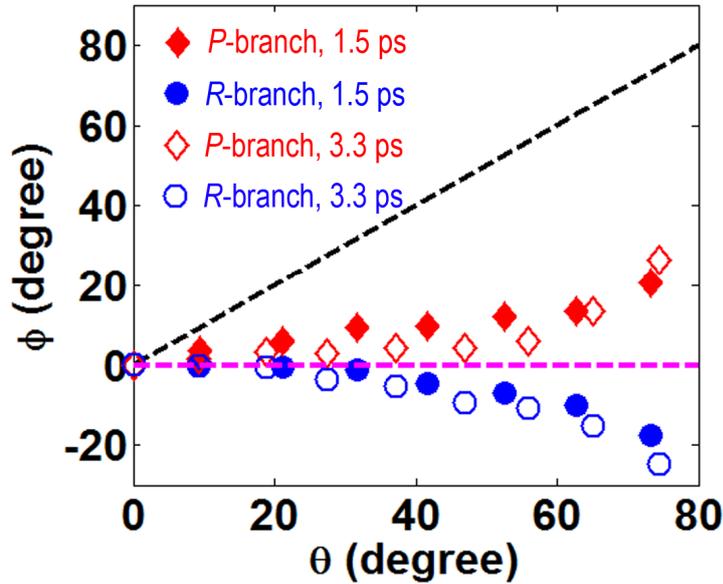

Fig. 3. The azimuthal angle $\phi$ of the P-branch (solid diamonds) and R-branch (solid circles) lines generated at the pump-probe time delay of 1.5 ps as the functions of the angle $\theta$. The azimuthal angle of the seed pulses (equivalent to the the angle $\theta$ here) measured in the wavelength range of 397 - 410 nm (i.e., the spectral range where the laser lines are excluded) as a function of the angle $\theta$ is presented with the black dashed line, showing a linear polarization unaffected by the aligned molecules. For comparison, same measurements on the azimuthal angles $\phi$ of the P-branch and R-branch lines generated at a pump-probe time delay of 3.3 ps are shown with the open diamonds and open circles, respectively.

To check how the polarization of the laser lines depends on the pump-probe delay, we changed the pump-probe time delay to 3.3 ps and performed the same measurements again. All the other parameters remain unchanged. As shown in Fig. 3 (see, the caption of Fig. 3), the data obtained at the two time delays almost overlap and only small quantitative difference is observed, indicating that the observed phenomenon is independent of the pump probe delay. We note that for both the time delays, the molecules are not at the revival times.

Figure 4(a) compares the azimuthal angles of the P-branch (diamonds) and R-branch (circles) laser lines as the functions of the angle $\theta$ for the $N_2^+$ lasers generated at gas pressures of 8 mbar (solid markers) and 27 mbar (open markers) with a pump pulse energy of 2 mJ. It can be seen that at both the pressures, the azimuthal angle of the P-branch laser lines increases with the angle $\theta$, whereas the azimuthal angle of the R-branch decreases with the angle $\theta$. Again, the qualitative feature obtained at the different gas pressures are similar to that in Fig. 3, whereas at the higher gas pressure, the polarization states of the P-branch and R-branch lines deviate more strongly from the linear polarization of the seed.

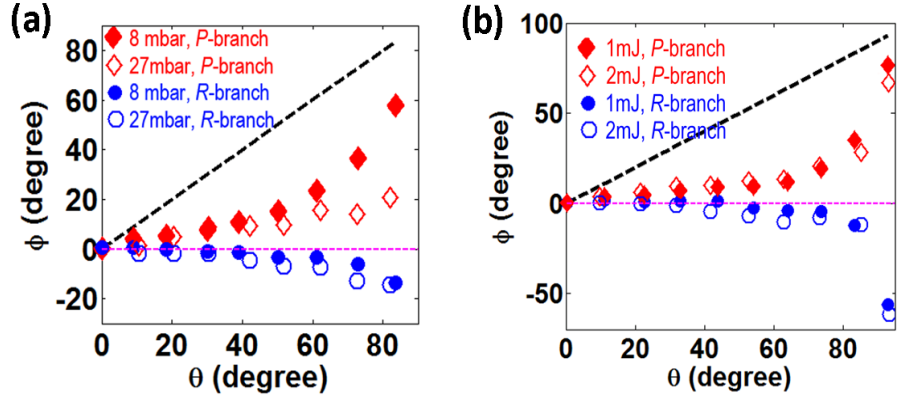

Fig. 4. The azimuthal angles $\phi$ of the *P*-branch (diamonds), *R*-branch (circles) laser lines and the seed pulses in the wavelength range of 397 - 410 nm (black dashed line) as the functions of the angle $\theta$ at (a) different gas pressures of 8 mbar (solid markers) and 27 mbar (open markers), and (b) different pulse energies of 1 mJ (solid markers) and 2 mJ (open markers).

At last, Fig. 4(b) compares the azimuthal angles of the *P*-branch (diamonds) and *R*-branch (circles) laser lines as the functions of the angle $\theta$ for the $N_2^+$ laser generated at the pump pulse energies of 1 mJ (solid marks) and 2 mJ (open marks) with a gas pressure of 13.5 mbar. The data measured at the two pump pulse energies almost overlap. It seems that the polarizations of both the *P*-branch and the *R*-branch laser lines are not sensitive to the pump pulse energy, probably due to the intensity clamping effect in the plasma channel which leads to stabilization of the peak intensity. We did not further increase the pump pulse energy, because self-seeded $N_2^+$ laser is generated when the pump energy is above 2 mJ. The self-seeded $N_2^+$ laser signal will contaminate the measurement of the dependence of polarization states of the $N_2^+$ lasers on the polarization of the external seed.

### 4. Discussion and conclusion

Typically, in the scenario of seed amplification, the laser signal should inherit the polarization property of the seed pulses. As a result, it is commonly expected that the generated $N_2^+$ laser should always be linearly polarized and the polarization direction should be parallel to that of the seed pulses. Surprisingly, this is not the situation as evidenced by our results given above.

A likely mechanism behind the unexpected observations in Figs. (2-4) could be that during the generation of the $N_2^+$ lasers through amplification of the seed pulses, the seed pulses inevitably experience strong birefringence. At the resonant wavelengths where the *P*-branch and *R*-branch laser lines are generated, abnormal birefringence exists which can be much stronger than the birefringence previously observed in molecular wavepackets aligned by femtosecond laser pulses [16]. In particular, for each individual laser line in the *P*-branch or *R*-branch emissions originated from the transition from a specific *J* state of the electronically excited molecules, the birefringence is only related to the molecular wavepackets associated to the rotational state *J*. Notably, the molecular wavepackets associated with the rotational state *J* will give rise to a strong anisotropy of the refractive index rapidly rotating in space, leading to the strong birefringence of both the *P*-branch and *R*-branch laser lines even at the delay times when no *overall* rotational revival occurs. Since the selection rules for the *P*-branch and *R*-branch laser lines are different, the two types of laser lines show ellipticities of different signs with the varying polarization direction of the seed pulses. The dependence of the azimuthal angles of laser lines on the gas pressure is a characteristic of the birefringence in aligned gaseous molecules. To confirm this mechanism, theoretical analyses based on sophisticated simulation tools should be carried out.

In conclusion, we investigated the polarization characteristics of the $N_2^+$ laser generated by two-color linearly polarized femtosecond laser pulses. We present an unexpected observation that the polarization of the $N_2^+$ laser does not always follow that of the seed pulses but can be controlled by changing the angle between the polarization directions of the pump and seed pulses. The *P*-branch and *R*-branch lines of the laser show different polarization characteristics as the angle between the polarization directions of the two pulses changes. Our finding not only reveals complex behaviors of propagation of ultrafast laser pulses in coherent rotational wave packets of molecules, but also enables generation of polarization-controllable free-space $N_2^+$ laser fields in air.